\documentstyle[epsfig]{mn}

\newif\ifAMStwofonts

\newcommand{\beq}{\begin{equation}}
\newcommand{\beqa}{\begin{eqnarray}}
\newcommand{\eeq}{\end{equation}}
\newcommand{\eeqa}{\end{eqnarray}}

\ifoldfss
  \ifCUPmtlplainloaded \else
    \NewTextAlphabet{textbfit} {cmbxti10} {}
    \NewTextAlphabet{textbfss} {cmssbx10} {}
    \NewMathAlphabet{mathbfit} {cmbxti10} {} 
    \NewMathAlphabet{mathbfss} {cmssbx10} {} 
  \fi
  \ifAMStwofonts
    \ifCUPmtlplainloaded \else
      \NewSymbolFont{upmath} {eurm10}
      \NewSymbolFont{AMSa} {msam10}
      \NewMathSymbol{\upi}     {0}{upmath}{19}
      \NewMathSymbol{\umu}     {0}{upmath}{16}
      \NewMathSymbol{\upartial}{0}{upmath}{40}
      \NewMathSymbol{\leqslant}{3}{AMSa}{36}
      \NewMathSymbol{\geqslant}{3}{AMSa}{3E}

      \let\leq=\leqslant 
      \let\geq=\geqslant 
    \fi
  \fi
\fi 

\ifnfssone
  \newmathalphabet{\mathit}
  \addtoversion{normal}{\mathit}{cmr}{m}{it}
  \addtoversion{bold}{\mathit}{cmr}{bx}{it}
  \newmathalphabet{\mathbfit} 
  \addtoversion{normal}{\mathbfit}{cmr}{bx}{it}
  \addtoversion{bold}{\mathbfit}{cmr}{bx}{it}
  \newmathalphabet{\mathbfss} 
  \addtoversion{normal}{\mathbfss}{cmss}{bx}{n}
  \addtoversion{bold}{\mathbfss}{cmss}{bx}{n}
  \ifAMStwofonts
    \ifCUPmtlplainloaded \else
      \UseAMStwoboldmath
      \makeatletter
      \new@mathgroup\upmath@group
      \define@mathgroup\mv@normal\upmath@group{eur}{m}{n}
      \define@mathgroup\mv@bold\upmath@group{eur}{b}{n}
      \edef\UPM{\hexnumber\upmath@group}
      \new@mathgroup\amsa@group
      \define@mathgroup\mv@normal\amsa@group{msa}{m}{n}
      \define@mathgroup\mv@bold\amsa@group{msa}{m}{n}
      \edef\AMSa{\hexnumber\amsa@group}
      \makeatother
      \mathchardef\upi="0\UPM19
      \mathchardef\umu="0\UPM16
      \mathchardef\upartial="0\UPM40
      \mathchardef\leqslant="3\AMSa36
      \mathchardef\geqslant="3\AMSa3E

      \let\leq=\leqslant 
      \let\geq=\geqslant 
    \fi
  \fi
\fi 

\ifnfsstwo
  \DeclareMathAlphabet{\mathbfit}{OT1}{cmr}{bx}{it}
  \SetMathAlphabet\mathbfit{bold}{OT1}{cmr}{bx}{it}
  \DeclareMathAlphabet{\mathbfss}{OT1}{cmss}{bx}{n}
  \SetMathAlphabet\mathbfss{bold}{OT1}{cmss}{bx}{n}
  \ifAMStwofonts
    \ifCUPmtlplainloaded \else
      \DeclareSymbolFont{UPM}{U}{eur}{m}{n}
      \SetSymbolFont{UPM}{bold}{U}{eur}{b}{n}
      \DeclareSymbolFont{AMSa}{U}{msa}{m}{n}
      \DeclareMathSymbol{\upi}{0}{UPM}{"19}
      \DeclareMathSymbol{\umu}{0}{UPM}{"16}
      \DeclareMathSymbol{\upartial}{0}{UPM}{"40}
      \DeclareMathSymbol{\leqslant}{3}{AMSa}{"36}
      \DeclareMathSymbol{\geqslant}{3}{AMSa}{"3E}

      \let\leq=\leqslant 
      \let\geq=\geqslant 
    \fi
  \fi
\fi 

\ifCUPmtlplainloaded \else
  \ifAMStwofonts \else 
    \def\upi{\pi}
    \def\umu{\mu}
    \def\upartial{\partial}
  \fi
\fi

\title{Detection of CO in the inner part of M\,31's bulge}
\author[A.-L. Melchior et al.] {
A.-L. Melchior,$^1$ F. Viallefond,$^2$ M. Gu\'elin,$^3$
N. Neininger$^4$\\
$^1$Astronomy Unit, Queen Mary and Westfield College, Mile End Road,
London E1\,4NS, UK\\ 
$^2$DEMIRM, Observatoire de  Paris, UMR8540, 61 Avenue de l'Observatoire,
75014 PARIS CEDEX, France\\ 
$^3$IRAM, 300 rue de la piscine, F-38406 St.Martin d'H\`eres, 
France\\ 
$^4$Radioastronomisches Institut der Universit\"at Bonn, 
Auf dem H\"ugel 71, D-53121 Bonn, Germany\\    
}
\date{accepted}

\pagerange{L\pageref{firstpage}--L\pageref{lastpage}}
\pubyear{1999}

\begin{document}

\maketitle			
\label{firstpage}

\begin{abstract} 
{We report the first detection of CO in M\,31's bulge. The
$^{12}$CO (1-0) and (2-1) lines are both detected in the dust complex
D395A/393/384, at 1$\farcm$3 ($\sim 0.35$ kpc) from the centre. From
these data and from visual extinction data, we derive a CO-luminosity
to reddening ratio (and a CO-luminosity to H$_2$ column density ratio)
quite similar to that observed in the local Galactic clouds. The (2-1)
to (1-0) line intensity ratio points to a CO rotational temperature
and a gas kinetic temperature $> 10$ K.  The molecular mass of the
complex, inside a 25$\arcsec$ (100 pc) region, is 1.5 10$^4$
$M_\odot$.}

\end{abstract}	

                 
\section{Introduction}
The bulk of gas of M\,31 {lies between 6 and 18 kpc and follows a
pattern of thin spiral arms (e.g. Dame et al.\ 1993, Neininger et
al. 1998) \nocite{Dame:1993, Neininger:1998}. Despite the presence of
dust lanes, massive stars, and evolved stars (e.g. O'Connell et
al. 1992; Davidge et al.\ 1997)\nocite{O'Connell:1992, Davidge:1997},
there is little HI atomic gas in the inner bulge and, up to now,
molecular gas had escaped detection: Sofue \& Yoshida
(1993)\nocite{Sofue:1993} reported the detection the $^{12}$CO(1-0) line in
the dust complex D395/393 at less than 1~kpc from the centre, but this
detection was invalidated by Loinard, Allen \& Lequeux
(1996)\nocite{Loinard:1996} who re-observed the same complex with a
much better sensitivity. So, the detection of CO in the dust complexes
D~478 and D~268, at 2-4~kpc from the centre\footnote{We assume a
distance to M\,31 of 780~kpc (e.g. Holland 1998; Stanek \& Garnavich
1998\nocite{Holland:1998,Stanek:1998c}), i.e.\ 1$\farcs$=3.8pc.}, by
Allen \& Lequeux (1993) and Loinard \& Allen
(1998)\nocite{Allen:1993,Loinard:1998}, was the closest to the centre
ever reported. Noting that the CO (1-0) to (2--1) intensity ratio was
lower in these clouds than in the Galactic Giant Molecular Clouds,
these authors concluded that CO may be so cold in the {\em inner disc}
of M\,31, that its excitation temperature barely exceeds the cosmic
background temperature. They thus suggested that although molecular
gas and CO could be abundant, the CO mm lines are so weak that they
escape detection.

In order to better understand the gas budget at the centre of M\,31,
we have embarked on a reanalysis of the optical data. We found that
the positions observed by Loinard, Allen \& Lequeux
(1996)\nocite{Loinard:1996} were not centred on the darkest dust
patches, which prompted us, taking advantage of the installation of
new generation mm-wave receivers on the IRAM 30-m telescope, to make
new CO observations.}

\section{Observations}  
\label{sect:obs}
\subsection{Distribution of the extinction in the bulge}
\label{ssect:ext}
We used a compilation of optical images of M\,31 {with resolutions
(FWHM) close to 1\,arc-sec to locate regions with significant
absorption in the bulge of M\,31:} (1) a $B$ frame ($0\farcs9$/pixel)
covering a field of $22\farcm9 \times 17\farcm9$ obtained during the
course of a 5 yr nova search by Ciardullo et al.\
(1988)\nocite{Ciardullo:1988}; (2) $B$ and $R$ images
($0\farcs3$/pixel), covering a field of $12'\times9'$, collected
during an intensive monitoring of M\,31's bulge aimed at micro-lensing
detections (Ansari et al.\ 1997,
1999)\nocite{Ansari:1997,Ansari:1999}, hereafter the AGAPE data; (3)
$BVRI$ images from J. Tonry covering a $4'\times4'$ field (Metzger,
Tonry \& Luppino\ 1993)\nocite{Metzger:1993}.  The astrometry
(M. Auri\`ere, private communication) of the optical data presents
small ($\sim 1$~arcsec) distortions on the degree scale; this is
insignificant given the size of the beam of the radio telescope
($21\arcsec$ at 115~GHz and $11\arcsec$ at 230~GHz).  We model the
photometry of M\,31's bulge with {elliptical annuli} using the
standard surface photometry algorithm developed for IRAF (Jedrzejewski
1987)\nocite{Jedrzejewski:1987}.  {This model intends to reproduce
the light profile along M\,31's bulge without extinction. Hence, the
ellipse geometry parameters (centre, ellipticity, position angle) are
fitted in this procedure.  In addition, to avoid possible
perturbations of the fit due to extincted areas, only the median
intensity over the elliptical annulus sectors is used.} The observed
extinction is {then} defined as A$_\lambda=-2.5\log_{10}(\phi_{\rm
obs}/\phi_{\rm model})$ where $\phi_{\rm obs}$ is the observed
brightness and $\phi_{\rm model}$ the brightness in the model. {
This derivation of A$_\lambda$, which assumes that the dusty cloud is
in front of the bulge, provides a lower bound of the true internal
extinction.}
\begin{figure*}
\resizebox{\hsize}{!}{\epsfig{file=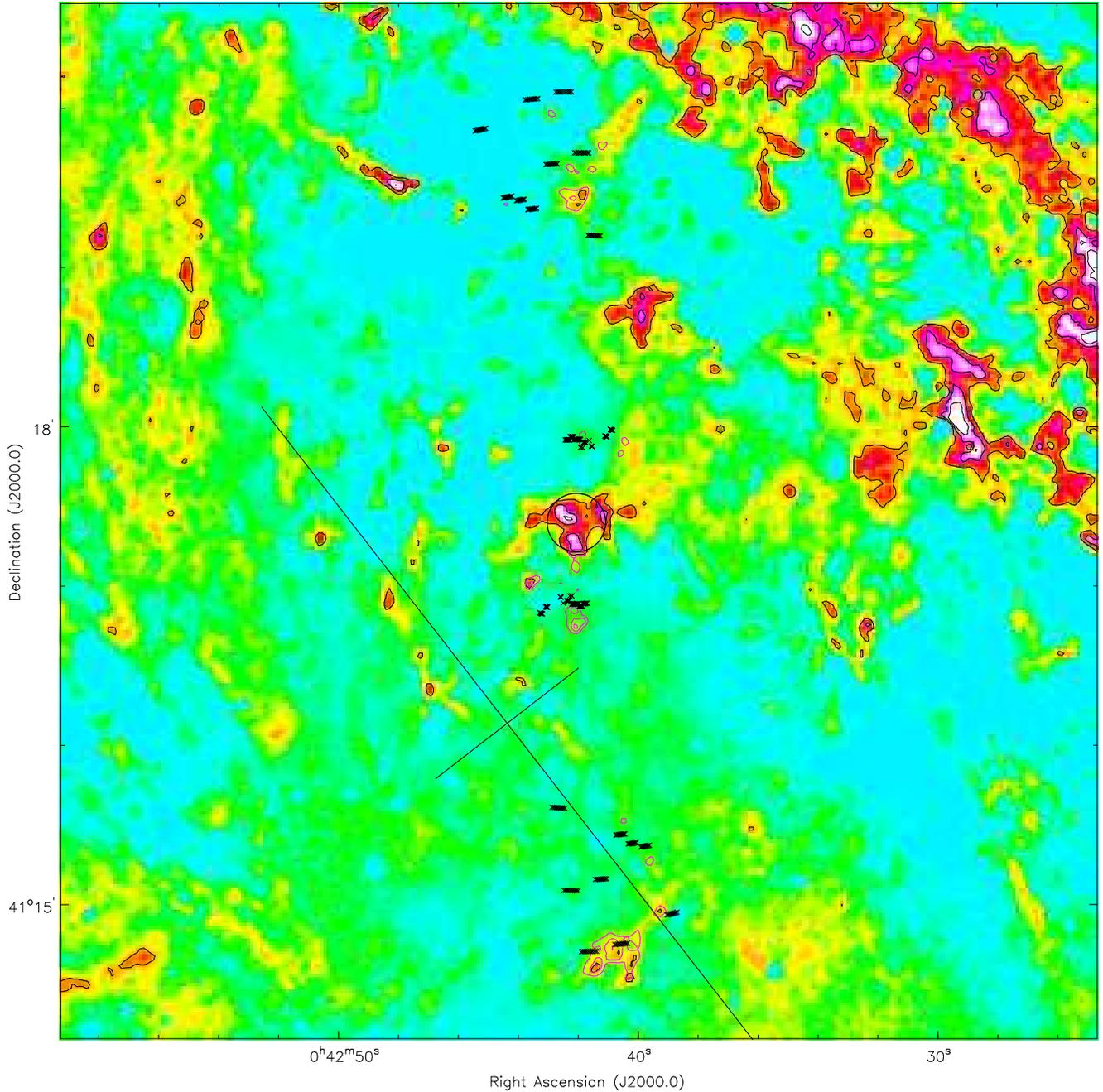,clip=,angle=-90}}
\caption{The A$_{B}$ extinction obtained from the data of Ciardullo et
al.\ (1988)\protect\nocite{Ciardullo:1988} is shown in {
grey}-scale. The black 
contour levels correspond to 0.1, 0.2 and 0.3~mag in A$_B$
scale. {The two straight segments} intersect at the centre of
M\,31. {Their lengths correspond} to 1~kpc along the major and minor
axis.  The circle shows the ON position toward the dust complex
D395A(/393/384). {Its diameter is the telescope} HPBW at 115~GHz.  The
{small} crosses display the OFF positions used {during} the
observations. The red 
contour levels give {the possible contamination level in the OFF
position at 115 GHz, assuming CO emission is proportional to visual
extinction} (first 0.05 per cent, second 0.10 per cent, third 0.15 per
cent).}
\label{fig:mapbig}
\end{figure*}

The comparison of 3 completely independent $B$ data sets shows that no
significant observational artifacts are introduced in the extinction
maps. The nominal centre of M\,31 is taken as the optical nucleus at
$\alpha_{J2000.0}=00^{\rm h}~42^{\rm m}~44^{\rm s}.371$ and
$\delta_{J2000.0}=41^{\circ}~16'~08\farcs34$ following Crane, Dickel
$\&$ Cowan\ (1992)\nocite{Crane:1992}.  Figure\,\ref{fig:mapbig} shows
the extinction obtained in $B$, centred on the dust patch
D395A/393. All the structures seen here are well correlated with those
seen in $B-R$. {We note} that the procedure defining the ellipses
tends to remove any smooth structures within 10~arcsec from the
centre. The structure in the north-western part is the edge of a much
larger extinction {pattern}; the most prominent structure detected
near the centre is D395A/393/384 with a typical size of 95~pc
($25$~arcsec). It is centred at $-26\arcsec$,$+76\arcsec$ from the
nucleus, {and consists of several dark patches or clumps.}

\subsection{Radio observations}
\begin{figure}
\resizebox{\hsize}{!}{\epsfig{file=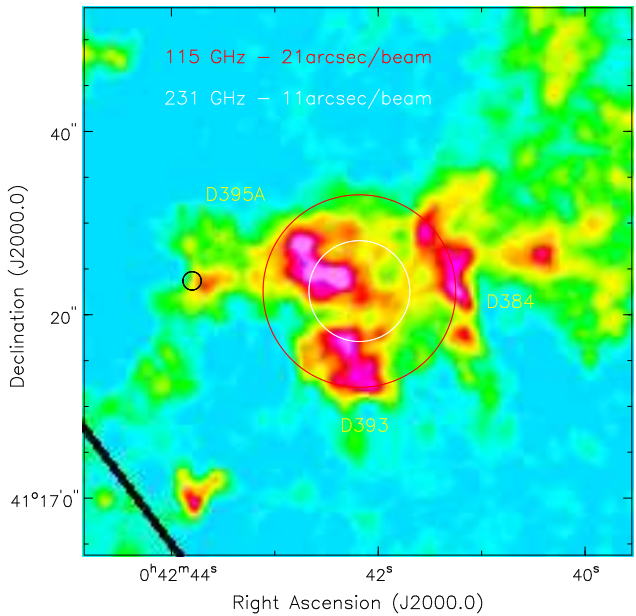,clip=,angle=-90}}
\caption{{The telescope HPBW at 115 GHz and 230 GHz during ON
scans} ($\alpha_{J2000.0}=00^{\rm h}42^{\rm m}42^{\rm s}.10$ and
$\delta_{J2000.0}=41^{\circ}17'24\farcs1$), superimposed on the {
A$_B$ extinction map derived} from the AGAPE data. The small circle
{ on the left marks the position observed} by Loinard et al.\
(1996).}
\label{fig:mapsmall}
\end{figure}
The observations were carried out on 1999 June 13-14 with the IRAM
30-m telescope.  Most of the observing was made in the symmetrical
wobbler switching mode { where} the secondary { mirror nutates}
up to a maximum limit of $\pm 240\arcsec$ in azimuth. { The beam
throw was determined as a function of the hour angle in such a way
that OFF positions lie in extinction-free regions (see
Fig.\,\ref{fig:mapbig}).} Near transit, we had to use position
switching mode, taking an extinction-free OFF position located at
$122\arcsec$,$321\arcsec$ from the nucleus.
Figure\,\ref{fig:mapsmall} show the areas covered by the telescope
beam { (HPBW) in} the ON position at 115 and 230~GHz. { The
beam} is centred as best as possible on the dust complex, and
encompasses most of D395A ({ and also, at 115 GHz, parts of D393
and D384}). The telescope pointing { was} checked every hour on {
nearby} quasars { and found to be} accurate within { 3~arcsec}.
\begin{figure}
\resizebox{\hsize}{!}{\includegraphics{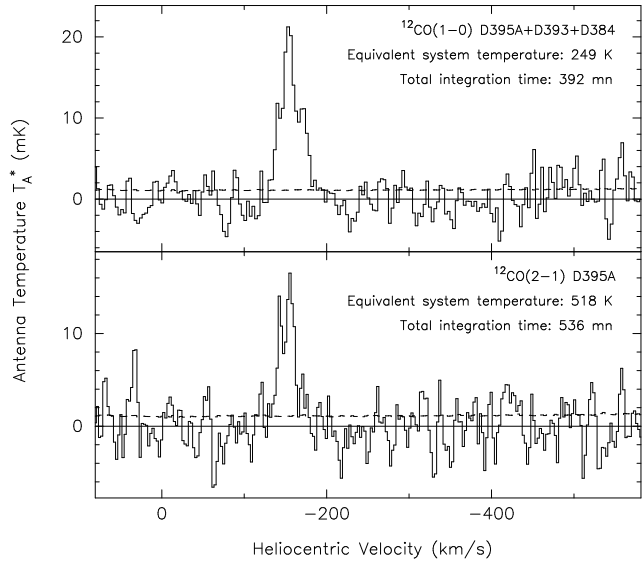}}
\caption{$^{12}$CO(1-0) and $^{12}$CO(2-1) spectra at a resolution of
3.0 km~s$^{-1}$, { observed towards} the dust complex
D395A(/393/384) (offsets from M\,31's nucleus:
$-26\arcsec$,$76\arcsec$).  The { r.m.s.} noise levels in
individual channels are illustrated by the dashed lines; they
correspond respectively to 1.8 and 2.9 mK. The integration times (IT)
quoted correspond to the sum of the ON and OFF times { for} two
receivers. IT is shorter for the (1-0) spectrum, because we excluded
the data with OFF positions closer than 2 HPBW ($50\arcsec$) from the
ON position.  The equivalent system temperature is { calculated
from the r.m.s. noise and IT, using the standard radiometer formula.}}
\label{fig:spectra}
\end{figure}

We used { four} receivers simultaneously, { two} for
$^{12}$CO(1-0) at 115~GHz and { two} for $^{12}$CO(2-1) at
230~GHz. { Each 115 GHz receiver} was connected to { two
autocorrelator sub-bands, shifted by 40 MHz from each other}. Each
sub-band consisted of 225 channels separated by 1.25~MHz. A very
simple algorithm, combining such a pair of settings, { was used to
remove occasional ``platforming'' between the 80 MHz wide units which
compose the sub-bands}. { Each 230 GHz receiver} was connected to a
filter-bank consisting of { 512 1-MHz-wide channels}. Linear
baselines were fitted and subtracted { from} the spectra registered
every 60~s, after the platforming correction for the data at
115~GHz. Some spectra were flagged, especially for one of the
receivers at 230~GHz, due to temporary instabilities.  The spectra
were then combined, {using inverse variance weights, which correspond}
to the residuals about the baselines.

The resulting spectra are displayed in Fig.\,\ref{fig:spectra} and the
lines parameters are summarised in Table\,\ref{tab:data}.  We estimate
that { even if the gas in} the OFF positions had the same velocity
{ as in the ON position} and if the CO luminosity was proportional
to A$_B$, we would have decreased at most by $6$ per cent the 115~GHz
line intensity (see the red 
contours in Fig.\,\ref{fig:mapbig}).  The { uncertainties} on the line
intensities quoted in Table\,\ref{tab:data} account for the { r.m.s.}
noise and for baseline uncertainties; { they do not include
calibration uncertainties} which, according to a check on the standard
source DR21, should be $\leq 10$ per cent. (We adopted a T$_A^*$ flux
scale corresponding to integrated $^{12}$CO(1-0) and $^{12}$CO(2-1)
emissions of respectively $410$ K km s$^{-1}$ and $360$ K km s$^{-1}$
at the centre of DR21 -- { see Mauersberger et
al. 1989}\nocite{Mauersberger:1989}). { We note that, although the
spectra of Fig.\,\ref{fig:spectra} were observed only 15$''$ away from
those of Loinard, Allen \& Lequeux (1996)\nocite{Loinard:1996}, the
line intensities are much weaker and the velocities completely
different from those reported by Sofue \& Yoshida
(1993)\nocite{Sofue:1993}. Hence, our detection is more in agreement
with the former authors, given their level of sensitivity.}

\begin{table}
\caption[ ]{Line parameters for the complex: $\langle v \rangle$ is
the heliocentric centroid velocity and $\sigma_{\rm FWHM}$ gives  the
measured FWHM of velocity profile. The CO flux, $S_{\rm CO}$,
ant the CO luminosity, $I_{\rm CO}$ (expressed in units of T$_A^*$,
the antenna temperature corrected for atmospheric absorption and rear
sidelobes) are given for both lines. $\eta_a$ is the the aperture
efficiency.}
\label{tab:data}
\begin{flushleft}
\begin{tabular}{lllcr}
$^{12}{\rm CO}$ & $\langle v \rangle$~~{ $\sigma_{\rm FWHM}$} & $I_{\rm
CO}$ & $S_{\rm CO}$ & $\eta_a$\\ \noalign{\smallskip} 
\multicolumn{5}{l}{\hspace{0.9cm} $\lgroup$km~s$^{-1}\rgroup$
\hspace{0.11cm} $\lgroup$K~km~s$^{-1}\rgroup$ \hspace{0.01cm}
$\lgroup$Jy~beam$^{-1}$\hfill}\\   
\multicolumn{5}{l}{\hspace{4.5cm}~km~s$^{-1}$ $\rgroup$\hfill}\\ \hline
\noalign{\smallskip}
(1-0) & -156~~$\sim$35 &  0.56$\pm
0.04$& 3.54 & ~~~~~~~0.57  \\ 
(2-1) & -154~~$\sim$23 &  0.36$\pm
0.05$& 3.81 & 0.32  \\ 
\end{tabular}
\end{flushleft}
\end{table}

\section{Discussion}
\label{sect:discu}
\subsection{Characteristics of the CO emission}
\label{ssect:char}
The spectra displayed in Fig.\,\ref{fig:spectra} show { that the
$^{12}$CO(1-0) and $^{12}$CO(2-1) emission is detected in the
heliocentric velocity range} $V_{\rm Helio}=-130,-190$ km~s$^{-1}$.
The measured centroid velocities are in good agreement with the
velocities of the ionised gas published by Boulesteix et al.\
(1987)\nocite{Boulesteix:1987}. This suggests that the molecular
clouds { lie} in the same plane as the ionised gas; the latter is
thought to be inclined by 45$^\circ$ in the region of interest (see
{ the discussion} by Ciardullo et al.\
1988\nocite{Ciardullo:1988}). { At the position observed, the
ionised gas presents a velocity gradient of $\sim$30~km$~$s$^{-1}$
over the 21$\arcsec$-wide telescope beam. This is consistent with
the FWHM of the CO(1-0) line profile (31~km$~$s$^{-1}$) measured for
the molecular complex.  This suggests that we are observing not one
single, but several clouds distributed across the telescope beam.}

The comparison of the intensities expressed as the $^{12}$CO(2-1) to
(1-0) line ratio R is not straightforward as the beams are different
and the geometry of the emitting regions is a priori unknown. If the
emitting region { were} uniform and much larger than the beams, the
line intensity ratio would simply be
R=T$_A^*$(2-1)/T$_A^*$(1-0)=0.6. We { adopt} the more realistic
assumption { suggested above that the dust follows the gas
distribution.}  Integrating the extinction distribution
(Fig.\,\ref{fig:mapsmall}) over the telescope beam and taking into
account the difference in telescope beam shapes and efficiencies
between 115 and 230~GHz (Greve, Kramer \& Wild
1998)\nocite{Greve:1998}, we arrive at a ratio R=0.5.  This value is
typical of those observed in M\,31's molecular arms (Neininger et al.\
1998)\nocite{Neininger:1998}
but is larger than those observed by Loinard et
al. (1996)\nocite{Loinard:1996}.

\subsection{The extinction curve for D395A/393/394}
\begin{figure}
\resizebox{\hsize}{!}{\includegraphics{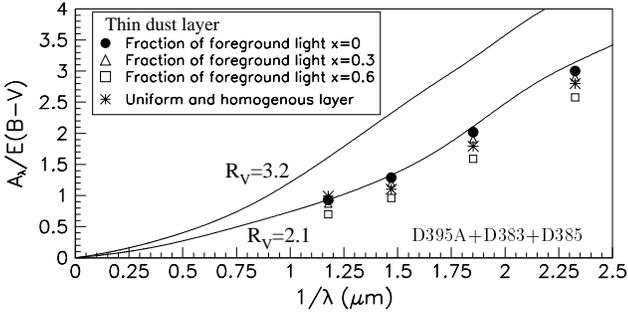}}
\caption{Extinction curve in the dust complex. Each symbol corresponds
to an average over $854$ pixels, with an extinction measured in $BVRI$
and for which A$_I > 0.04$. They show the extinction law estimated for
different geometries of the dust. The lines correspond to the models
of Fitzpatrick (1999)\protect\nocite{Fitzpatrick:1999}.}
\label{fig:ext}
\end{figure}
The extinction curve is determined by comparing the extinction in $B$,
$V$, $R$ and $I$ on a pixel basis. Figure\,\ref{fig:ext} shows that
the A$_V$/$E(B-V)$ ratio R$_V$ is closer to 2.0 than to the canonical
value of 3.1. This cannot be accounted by the geometry of the dust
distribution (see Walterbos \& Kennicutt (1988) for a comprehensive
review)\nocite{Walterbos:1988}. If dust and stars are uniformly mixed
along the line of sight, the true extinction would then be increased
by a factor of 2, but the extinction curve would still present the
same R$_V$. If the dust is located in a thin layer, which is rather
plausible if associated with the molecular and ionised gases, the
observed extinction, as defined in Sect. \ref{ssect:ext}, then
corresponds to $-2.5\log_{10}(x+(1-x)\times \exp(-\tau))$, where
$\tau$ is the real optical depth at a given wavelength and $x$ the
fraction of light in front of the dust. As observed extinctions as
large as 0.49~mag are measured in $B$, the fraction of foreground
light $x$ has to be smaller than 0.6. This fraction is 0.3 and 0.02 if
the dust patch lies in a plane inclined by 45$^{\circ}$ and
77$^{\circ}$ respectively, when using Galactic bulge and disc models
from Binney, Gerhard \& Spergel\ (1997)\nocite{Binney:1997}. Similar
deviations from the average R$_{V}$ have already been reported in a
few lines of sight across M\,31's disc (Massey et al. 1995, and
references therein)\nocite{Massey:1995}: like what is observed in the
Galaxy (e.g. Szomoru \& Guhathakurta 1999\nocite{Szomoru:1999}), local
fluctuations are also present in M\,31.

For $x=0$, we estimate the extinction A$_V$=0.06 mag beam$^{-1}$ when
averaged over the 115~GHz beam, corresponding to $E(B-V)$=0.0285 mag
beam$^{-1}$. Given the arguments discussed above, the uncertainties
affecting these values are negligible for the following.

\subsection{The gas-dust connection}
Although it is very dubious to interpret the observed CO luminosity in
terms of mass of molecular hydrogen, the I$_{\rm CO}$/$E(B-V)$=9.8
K~km~s$^{-1}$~mag$^{-1}$ ratio measured here is remarkably similar to
the value observed for the molecular clouds in our Galaxy. Hence, the
most direct method to get the order of magnitude for the mass of the
gas in this complex is to assume a standard dust-to-gas mass ratio,
N$_{\rm H}$/$E(B-V)$ = 5.8 10$^{21}$ atoms cm$^{-2}$ mag$^{-1}$
(Bohlin, Savage $\&$ Drake, 1978)\nocite{Bohlin:1978}. Integrating
directly over the distribution of extinction, we derive a mass of
1.5~$10^4$~M$_\odot$ (10$^4$~M$_\odot$ in the main solid angle of the
beam at 115~GHz). This mass could be { overestimated if the
metallicity is larger than 1 in the central region of
M\,31}. Continuing, based on the optical extinction map, we find that
the size of this complex is about 60~pc; this is typical for a giant
molecular complex. However, its average volume density ($\sim
1$~cm$^{-3}$) is extremely low: two orders of magnitude lower than the
typical densities in giant molecular clouds.  For this reason, this
complex must be highly clumped as supported by the detection of these
CO lines.

If these clumps are gravitationally bound together, we estimate a
virial mass { $M_{\rm vir}/$M$_\odot \sim 200 r/{\rm (pc)}
(\sigma_{\rm FWHM}/{\rm (km/s)})^2$, i.e. $M_{\rm vir} \sim
1.6~10^7$~M$_\odot$}. The difference of { three} orders of magnitude
compared to the above result based on the optical extinction is so
large that either these clumps are not bound together or the mass is
not dominated by the gaseous component. For the very same reason, this
complex does not follow the velocity line-width versus diameter
relationship found by Solomon et al.\ (1987)\nocite{Solomon:1987} for
Galactic molecular clouds near virial equilibrium, but presents
similar characteristics to the clouds studied by Oka et al.\
(1998)\nocite{Oka:1998} in the Galactic centre.

\section{Conclusions}
Emission of the (1-0) and (2-1) { lines} of the $^{12}$CO molecule
has been detected near the centre of M\,31 at $15\sigma$ and
$10\sigma$ respectively. The CO centroid velocity of
$-155$~km~s$^{-1}$ suggests that this molecular gas could be located
in the ionised gas disc detected in these regions. { The observed
velocity dispersion of the molecular gas in the beam is compatible
with the velocity gradient of the ionised gas. Together with the
patchy appearance of the extinction map, this indicates the presence
of several clouds distributed over the beam.}  The I$_{\rm
CO}$/$E(B-V)$ ratio is remarkably similar to the values observed in
molecular clouds in the Galaxy and M\,31's disc.  Based on a standard
gas-to-dust ratio, the mass of the molecular complex is of order
10$^4$~M$_{\odot}$. { In order to reach densities compatible with
the excitation of the CO lines,} the complex must be highly clumped.
As previous studies with poorer resolution have shown, the mass
inferred is small compared to the mass content of the Galactic centre
(Morris \& Serabyn 1996)\nocite{Morris:1996}. With such a mass the
line-width of the CO line is by far too broad for a gravitationally
bound complex, which explains previous inconsistencies based on this
assumption.

{ Accordingly, we modelled the extinction
with a random distribution of small spherical clumps, all identical
with a size $r$ and density $\rho$. They all lie in a sphere of radius
13$\arcsec$, tracing a total mass of $10^4$ M$_\odot$. Models with
$\sim$250 clumps reproduce the main features of the histogram of
measured extinctions. With the previous assumption of a common
location for the molecular, ionised gas and the dust, the
configuration (r,$\rho$) which reproduces best the measured extinction
correspond to clumps with r$\sim$2.6$\arcsec$ and
$\rho\sim$400~H~cm$^{-3}$. Following the LVG homogeneous cloud models
of Garc\'\i a-Burillo, Gu\'elin $\&$ Cernicharo
(1993)\nocite{Garcia-Burillo:1993}, the measured line ratio
corresponds to CO (1-0) and (2-1) line excitation temperatures and
to a kinetic temperature all $\geq$10~K. This modelling will be
further discussed in a subsequent paper. 

The ``standard'' CO-luminosity to visual extinction ratio and the
relatively high CO (1-0) excitation temperature seem to rule out the
presence of large amounts of hidden CO and H$_2$ in the inner bulge of
M\,31, in apparent disagreement with the result of Loinard et
al. (1996)\nocite{Loinard:1996} in the inner disc.}

\section*{Acknowledgements}
We thank the AGAPE collaboration, R. Ciardullo, G. Jacoby, E. Magnier
and J. Tonry for providing their optical data. We are grateful to
J. Lequeux for valuable comments. It is a pleasure to thank Richard
Frewin for his help with various software issues. During this work,
A.-L. Melchior has been supported by a European contract
ERBFMBICT972375 at QMW.

\label{lastpage}

\end{document}